\documentclass[fleqn,usenatbib]{mnras}

\usepackage{newtxtext,newtxmath}

\usepackage[T1]{fontenc}
\usepackage{anyfontsize}

\DeclareRobustCommand{\VAN}[3]{#2}
\let\VANthebibliography\thebibliography
\def\thebibliography{\DeclareRobustCommand{\VAN}[3]{##3}\VANthebibliography}

\usepackage{graphicx}	
\usepackage{amsmath}	
\usepackage{bm}

\title[\bl\ dynamo in rapidly rotating stars]
{Does the Babcock–Leighton dynamo operate in rapidly rotating solar-type stars? Exploration using a 3D dynamo model at different rotation rates}

\author[V. Vashishth et al.]{
Vindya Vashishth,$^{1}$\thanks{E-mail: vindyavashishth.rs.phy19@itbhu.ac.in (VV)}
Bidya Binay Karak,$^{2}$\thanks{E-mail: karak.phy@iitbhu.ac.in (BBK)}
\\
$^{1,2}$Department of Physics,  
Indian Institute of Technology (BHU), 
Varanasi 221005, India
\\
}

\date{Accepted XXX. Received YYY; in original form ZZZ}

\pubyear{\the\year{}}

\newcommand{\Fig}[1]{Figure~\ref{#1}}

\newcommand{\Eq}[1]{Equation~(\ref{#1})}

\newcommand{\Sec}[1]{Section~\ref{#1}}

\def\bl{Babcock--Leighton}

\def\dr{differential rotation}
\def\Rs{R_{s}}

\newcommand{\etas}{\eta_{\mathrm{S}}}
\newcommand{\etaCZ}{\eta_{\mathrm{CZ}}}
\newcommand{\etaRZ}{\eta_{\mathrm{RZ}}}
\newcommand{\rBCZ}{r_{\mathrm{BCZ}}}
\newcommand{\rsurf}{r_{\mathrm{surf}}}

\begin{document}
\label{firstpage}
\pagerange{\pageref{firstpage}--\pageref{lastpage}}
\maketitle

\begin{abstract}
The Babcock-Leighton dynamo, which relies on the generation of a poloidal field through the decay and dispersal of tilted bipolar magnetic regions (BMRs), is a promising paradigm for explaining the features of the solar magnetic cycle. In rapidly rotating stars, BMRs are expected to emerge at high latitudes, which are less efficient in generating the poloidal field due to poor cross-equatorial cancellation. The operation of the Babcock-Leighton dynamo in rapidly rotating stars is therefore questionable. We, for the first time, using a 3D kinematic dynamo model, STABLE, explore this question. By taking large-scale flows from mean-field hydrodynamics models for stars rotating at different speeds, We conduct a series of dynamo simulations in rapidly rotating stars, exploring the following four cases of spot deposition, each based on a different assumption about toroidal flux tube rise: (i) radial rise, (ii) parallel rise to the rotation axis, (iii) parallel rise combined with an increase in Joy's law slope with the stellar rotation rate, and (iv) increasing time delay and spot size.
We find cyclic magnetic fields in all cases except case IV of the 1-day rotating star, for which the magnetic field is irregular. For the parallel-rise cases, the magnetic field becomes quadrupolar, and for all other cases, it is dipolar.
Our work demonstrates that the Babcock-Leighton dynamo may operate even in rapidly rotating stars with starspots appearing at higher latitudes.

\end{abstract}




\section{Introduction}

\label{sec:intro}

Low main sequence stars exhibit magnetic cycles analogous to the 11-year solar cycle \citep{Baliu95}. 
These cycles are characterized by the periodic variation in the star's magnetic field, manifested through 
e.g., changes in starspot number, chromospheric activity, and coronal emissions. The properties of these cycles, such as duration, amplitude, and overall activity, vary significantly among different stars. 
These variations are influenced by factors such as the star's age, rotation rate, surface temperature, internal structure, etc.
In general, the more rapidly a star rotates, the more active it gets \citep{Skumanich72, R84}. \citet{Noyes84a} and \citet{WD16} gave the activity-rotation relation using Ca II H \& K and X-ray emissions, respectively. They demonstrated that activity increases with the rotation rate for slow rotators but tends to saturate for fast-rotating stars. There are a few pieces of evidence that suggest the activity may even decline somewhat in the most rapid rotators 
\cite[e.g.,][]{James2000}. A similar resemblance is also observed between rotation rate and estimates of the unsigned surface magnetic flux \citep{Saar96,Saar01, Reiners09}.
A complete theoretical understanding of the rotation-activity relationship is still not clear \citep{isik2023}. These phenomena likely depend on factors like magnetic flux emergence, chromospheric and coronal heating, and mass loss mechanisms. However, it is widely believed that the basic rotation-activity relationship is due to an inherent connection between rotation and the dynamo process \citep{Noyes84a, Baliu95}.

A large-scale dynamo, driven by the helical convection and \dr, is fundamental to generating and sustaining the magnetic fields in the Sun \citep{Pa55}. 
As the sun-like stars have convection zones (CZs) in their outer layers like the Sun, it is natural to expect that these stars also support dynamo action through which the stellar magnetic cycles are maintained. Some of the stars (e.g., HD 10476, HD 16160, etc.) have cycles similar to the solar cycle, which suggests that a similar dynamo that is operating in our Sun might be working in other sun-like stars \cite[also see][]{garg19, jeffers22, Jeffers2023}.
In this dynamo process, shearing due to differential rotation produces a toroidal magnetic field from a poloidal one (the $\Omega$ effect), while cyclonic convection (the $\alpha$ effect) regenerates the poloidal magnetic field.
Although the $\alpha$ effect, which involves the lifting and twisting of the toroidal field due to helical convection, is a potential way to generate the poloidal field, the Babcock-Leighton (BL) process has strong observational support and is now considered the main method for generating the poloidal field in the Sun \citep[e.g.,][]{Das10, KO11b,Priy14, CS15}.
In this process, tilted sunspot groups (more accurately, the bipolar magnetic regions) decay and disperse to produce a poloidal field through turbulent diffusion, meridional flow, and differential rotation. 

The key ingredients in the BL process are the tilt and the latitudinal position of BMR emergence. In fact, surface flux transport (SFT) simulations \citep{JCS14}
dynamo models \citep{KM18, Kumar24}, and analytic calculations \citep{Pet20} 
show that the higher the latitude, the less effective the poloidal field generation. This phenomenon is a key to the so-called latitudinal quenching \citep{Petrovay20, J20, Kar20, yeates25, Dey25}, implying that stronger cycles that produce sunspots at higher latitudes are less effective in generating a poloidal field. In short, latitude plays a crucial role in the effectiveness of the BL process. 

Since star spots are also expected to show similar features as of sunspots, we expect BL process to operate in other stars. Previous studies already employed axisymmetric kinematic dynamo model by parameterizing BL process with a non-local $\alpha$ effect to explain features of stellar cycles \citep{NM07, JBB10, KKC14,  Hazra19, Vindya23}. 
In these previous stellar dynamo models, the BL process is not adequately captured; it is through a non-local $\alpha$ term which is included as a poloidal field source term. 
Star spots are believed to be produced through the rise of the toroidal flux tubes from the deeper CZ and during the rise of flux tubes, the Coriolis force acting on the diverging flows arising from the apex of the tubes causes the tilt \citep{DC93, sreedevi25}. Thus, we expect the tilt angle and the emerging latitude to increase with the increase of rotation rates of stars \citep{Schu1992, isik2018}. 
Most stars are born 
with rapid rotation, and in their early stages, 
the flows in their CZs 
experience more Coriolis force, leading to larger tilt angles and spots appearing at high latitudes. Observations also suggest that in young-rapidly rotating sun-like stars, spots appear at high latitudes \citep{Schu1996, Luo22}. This naturally raises a question: Does the BL process operate in rapidly rotating stars?

In the present study, we employ a 3D kinematic solar dynamo model STABLE \citep[Surface flux Transport And Babcock–LEighton;][]{MD14, MT16} to explore the functioning of the BL process in the stars with rotation rates varying from 1 day to 30 days. 
In this work, we incorporate meridional flow and differential rotation from a mean-field hydrodynamics model \citep{KO11} for stars of different rotation periods and depths of CZ. Our results have important implications for magnetically active stars, particularly young, fast rotators, whose magnetic cycles exhibit significant variability in strength and duration \citep[e.g.,][]{Baliu95, garg25}.
Thus, our study aims to identify how the BL process operates across a range of stars, from rapid to slow rotators, and how cycle strength and duration vary with rotation rate.
In \Sec{sec:models}, we present our model, while in \Sec{sec:results} we discuss our results. Finally, in \Sec{sec:conclusion}, we summarize our results and highlight the conclusion.

\section{STABLE model}\label{sec:models}
We build on our study by using a 3D kinematic dynamo model,  STABLE \citep{MD14, MT16, KM17, HM16, Kinfe24}, 
which solves the following induction equation in three dimensions encompassing solar CZ. 
\begin{equation}
\frac{\partial \bm B}{\partial t} = \nabla \times \left[  \bm v \times \bm B - \eta_t \nabla \times \bm B \right],
\label{eq:induction}
\end{equation}
where the large-scale velocity, $\bm V$ is represented as,
\begin{equation}
{\bm v} = {\bm v_p} + r \sin\theta~\Omega (r,\theta) { \hat \phi}.
\end{equation}
Here,  ${\bm v_p} =  v_r^\prime { \hat r} + v_{\theta} {\bf \hat \theta}  =  (v_r +\gamma_r) { \hat r} + v_{\theta} {\bf \hat \theta}$, which includes the meridional circulation ($v_r,v_\theta$) and the radial pumping ($\gamma$), $\eta_t$ is the effective turbulent diffusivity that incorporates the mixing effect of the small-scale convective flow, and $\Omega$ is the angular velocity.

The profile of the radial magnetic pumping is the same as shown by \citet{KC16}. This radial pumping is needed to make the dynamo model in agreement with the surface flux transport model \citep{Ca12}. It also suppresses the diffusion of the magnetic field through the surface and thus helps the model to produce solar-like 11-year cycles at high diffusivity value \citep{KM17}, which otherwise becomes too short \citep{KC12}. The latitudinal component of the magnetic pumping, which is also possible in the rotating convection zone but not considering in the present study, helps to transport the toroidal magnetic field at the base of the convection zone, which causes equatorward migration of sunspots without the meridional flow \citep{Hazra_2016}.
All the simulations performed in this paper include downward radial magnetic pumping. If magnetic pumping is not included, the dynamo decays—consistent with what is also found in solar dynamo models.

For $\eta_t$, we take it as a function of $r$ alone and has the following form:
\begin{eqnarray}
\eta_t(r) = \etaRZ + \frac{\etaCZ}{2}\left[1 + \mathrm{erf} \left(\frac{r - \rBCZ}
{d_1}\right) \right]\nonumber \\
+\frac{\etas}{2}\left[1 + \mathrm{erf} \left(\frac{r - \rsurf}
{d_2}\right) \right],
\label{eq:eta}
\end{eqnarray}\\
with $\rBCZ=0.715 \Rs$, $d_1=0.0125 \Rs$, $d_2=0.025 \Rs$, $\rsurf = 0.956 \Rs$, $\etaRZ = 1.0 \times 10^9$ cm$^2$~s$^{-1}$, $\etaCZ = 1.5 \times 10^{12}$ cm$^2$~s$^{-1}$, and $\etas = 3 \times10^{12}$ cm$^2$ s$^{-1}$.

The SpotMaker algorithm is a key component of the STABLE model. This algorithm places Bipolar Magnetic Regions (BMRs) on the stellar surface based on the underlying azimuthal (toroidal) magnetic field 
at the base of the CZ, 
computed as:
\begin{equation}
\hat{B}_\phi(\theta, \phi, t) = \int_{r_a}^{r_b} h(r) B_{\phi}(r, \theta, \phi, t)dr, 
\label{eq:Bhat}
\end{equation}
where $r_a$ \&  $r_a$ are $0.70 R_s$ \& $0.715 R_s$, respectively, and $h(r) = h_0(r - r_a)(r_b - r)$ with $h_0$ being the normalization factor.

When this field exceeds an assigned threshold value ($B_t = 2000$~G) and the time delay between the two successive spots is greater than $\Delta$, then this algorithm adds a spot on the surface at the same latitude.  
The time delay $\Delta$
is taken from the following log-normal distribution of the time delay of BMRs (consistent with solar observations; \citet{Kumar24}),
\begin{equation}
N(\Delta) = \frac{1}{\sigma_d \Delta \sqrt{2\pi}} \exp \left[ - \frac{(\ln\Delta - \mu_d)^2}{2\sigma_d ^2}\right],
\end{equation}
where, $\sigma_d ^2 = (2/3)[\ln \tau_s - \ln \tau_p]$, with $\tau_s$ and $\tau_p$ as mean time between two consecutive spots and mode of the distribution respectively, and  $\mu_d = \sigma_d ^2 + \ln \tau_p$. 
$\tau_p$ and $\tau_s$ are expressed in terms of the azimuthal-averaged toroidal magnetic field ($B_b^N$) in a thin layer from $r = 0.715 R_s$ to $0.73 R_s$ around $15^\circ$ latitudes and $B_{\tau}$ = 400 G, and is given as;
\begin{equation}
\tau_p = \frac{2.2~\rm{days}}{1+ (B_b^N/B_{\tau})^2},~~
\tau_s = \frac{20~\rm{days}}{1+ (B_b^N/B_{\tau})^2}.
\label{eq:delay}
\end{equation}
Now, when $\hat{B} > B_t$ and $dt > \Delta$, then the SpotMaker adds a BMR on the surface. Note that these conditions are independently checked in two hemispheres, and no hemispheric symmetry is imposed. 
Once the timing of the spot is decided, the flux of the spots is taken from their observed distributions (Equations 8 of \citet{KM17}). 

For the tilt of the BMR, we use the standard Joy's law, i.e., 
\begin{equation}
\delta = \frac{\delta_0 \cos \theta}{1 + (B / B_{\rm sat})^2}
\label{eq:Joy}
\end{equation}
where $\delta_0 = 35^\circ$, 
$\theta$ is colatitude. To limit the growth of the magnetic field in dynamo, a magnetic field-dependent quenching in the tilt angle (of the form $1 / (1 + (B / B_{\rm sat})^2)$, where $B$ is the average $B_\phi$ at BCZ and $B_{\rm sat} = 100$~kG) is included as inspired by observations \citep{Jha20, sreedevi24}.

\section{Application of STABLE to stars}
The STABLE dynamo model was originally designed to reproduce the magnetic field of the Sun by utilizing the large-scale flows and the deposition and decay of BMRs on the solar surface (BL process) guided by solar observations. This model produces many basic features of the solar cycle, including its long-term variation \citep{KM17, KM18, Kar20, Mord22}. However, when we apply this model to stars, we need to make some essential modifications to it. The obvious modification will be in the large-scale flow.

We take the profiles of the meridional circulation and the differential rotation from the mean-field hydrodynamics models of \citet{KO11} for stars of rotation periods of 1, 3, 7, 10, 15, 20, 25.38 (solar value), and 30 days. 
\citet{KO11} model produces solar-like differential rotations for all these stars by solving the steady state equation of motion and the entropy equation in combination of EZ stellar evolution code \citep{Paxton} to specify the structure of a $1 M_0$ mass star as a function of age and the gyrochronology relation to identify the rotation rate for the star of a given age \citep{barnes05}. 
The differential rotation obtained from this model for the Sun matches closely with helioseismic observations and for the rapidly rotating stars, it also agrees with the surface differential rotation \citep{barnes05}.
The profiles for the differential rotations are given in several prior publications, notably outlined in \citet{KKC14, Hazra19} and thus not reproduced again here.

With the above flows, we conduct several stellar dynamo simulations to explore the operation of BL process in solar-type stars. We consider following cases. 

\subsection{Case I}
In this case, we do not make any changes in the STABLE model except the large-scale flow as mentioned above. We recall that in this model, the SpotMaker algorithm places a spot on the surface at the same latitude where it is linked to the azimuthal field at the base of CZ; that is, if $\theta_b$ is the latitude of the BMR where  $\hat{B}$ (as computed from \Eq{eq:Bhat}) exceeds $B_t$ in the CZ, then $\theta_s$ is the latitude where the spot is deposited on the surface, then we have taken, $\theta_b = \theta_s$. This scenario is referred to as Case I in this work, where the spot is positioned radially.

\subsection{Case II}

Observations show spots near the poles in rapidly rotating stars \citep[e.g.,][]{Stra1996}. 
The possible explanation for these polar spots is the stronger effect of the Coriolis force on the toroidal flux tube, which causes the flux tubes to rise parallel to the rotation axis as they travel from the deeper CZ to the surface \citep{Schu1992, Schu1996, Luo22}.
Some observations also suggest that a few rapidly rotating stars can show spot emergence at low latitudes \citep{Stras98, barnes05}, however, high-latitude (or polar) spots remain the dominant feature.  For the sake of simplicity, we therefore focus on high-latitude spots in this study.

Theory shows that for a rotation period less than about 10 days, the toroidal flux tubes rise parallel to the rotation axis  \citep{Schu1992}.
Therefore, in this case, we capture this effect 
by placing the spots parallel to the axis of rotation for stars of rotation period shorter than 10 days, while maintaining a solar-like tilt angle based on Joy’s law \Eq{eq:Joy}.Accordingly, $\theta_s$
will be computed based on the value of $\theta_b$ to place the spot parallel to the rotation axis.
This scenario, where everything else is the same as Case~I, is referred to as Case~II.

\subsection{Case III}
We move further by capturing the increase of the tilt angle with the rotation rate of stars. So far in Cases I-II, we were considering the amplitude of Joy's law the same for all stars and it was given by \Eq{eq:Joy}. However, theoretical studies using thin flux tube model simulations indicate that the tilt angle of BMR increases with the rotation rate \citep{isik2018}. 
Hence, in this case, we make the tilt rotation dependent in addition to the parallel rise of the toroidal flux for stars with rotation period less than 10 days (Case~II).
Since the relationship between tilt and rotation rate is not well understood, we use the following form to capture this dependency:
\begin{equation}
\delta_0 =  \delta_0 \left( \frac{P_{s}}{P_{\ast}}\right)^{\zeta},
\label{eq:joyscalling}
\end{equation}
where $P_{\ast}$ and $P_{s}$ are the rotation period of the star and the sun, respectively, and $\zeta$ is the factor by which  $\frac{P_{\ast}}{P_{s}}$ influences the tilt angle.

\subsection{Case IV}

Finally, we apprehend the timedelay and the size of the starspots in rapidly rotating stars. Several observations suggest that fast rotators produce big starspots at higher latitudes, which can persist for several years as well \citep{Strassmeier1999,hall+HENRY1994,Roettenbacher2017}. One can assume that the timedelay between the observed starspots is much longer in rapidly rotating stars, allowing for the formation of these large spots. 
To incorporate this effect, we consider two new modifications:

(a) We increase the time delay between starspots. For this, we update the mean time between spots and the mode of the distribution, $\tau_p'$ and $\tau_s'$ as follows, 
\begin{equation}
\tau_p' =  \left(0.4 + \tau_p \right)\left( \frac{P_{s}}{P_{\ast}}\right), \tau_s' =  \left(1.9 + \tau_s \right) \left( \frac{P_{s}}{P_{\ast}}\right).
\end{equation}

(b) And we increase the area of the starspots. To achieve this, we scale the BMR flux distribution linearly with the toroidal field at the base of the CZ, and can be represented as, 
\begin{equation}
\Phi_s = \left( B(\theta_s, \phi_s,t)/B_{\rm{sat}} \right)\Phi.
\label{eq:flux}
\end{equation}
Here, $(\theta_s,\phi_s)$ denotes the location of the spot, and $\Phi$ can be obtained using the observed distribution of spot flux \citep{Mu15}.


\section{Results and Discussion} \label{sec:results}
We have performed dynamo simulations for $1~M_\odot$ stars with rotation periods ranging from 1 day to 30 days for different cases as discussed above. 
Since the primary objective of this work is to explore dynamo operation in rapidly rotating stars, we first present a detailed analysis of the results for the star with a 1-day rotation period.

\subsection{Star of 1-day rotation period}
\begin{figure}
    \centering
    \includegraphics[width=\columnwidth]{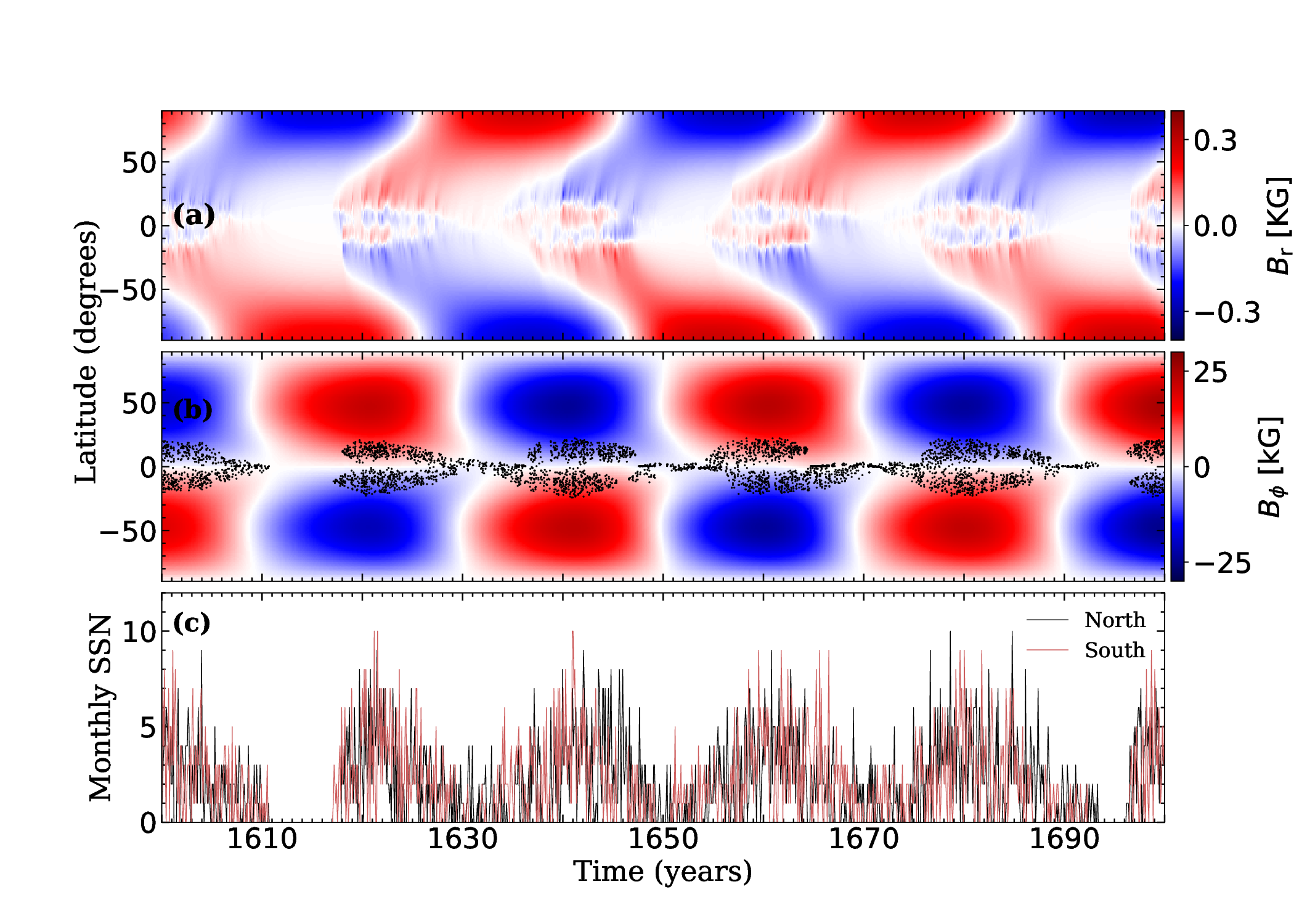}
    \caption{(a) Time-latitude distribution of the surface radial magnetic field $B_{\rm r}$ [in kG], and (b) toroidal field along with starspot distribution (black dots) for a star of 1 day rotation period for Case I.
    (c) Monthly number of spots as function of time. 
    } 
    \label{fig:rot1r}
\end{figure}

The star with a rotation period of 1 day produces regular polarity reversals approximately every 20 years in Case I, as shown in \Fig{fig:rot1r}. The magnetic field is predominantly dipolar, with a strong toroidal field concentrated around 50° latitudes.
In this case (radial spot deposition), spots are deposited at low latitudes, which maintain a dipolar magnetic field through the efficient cross-equatorial cancellation of the leading polarity flux from the other hemisphere \citep[e.g.,][]{Durrant04, Ca13, KM18}. The results drastically change in Case II, where spots are deposited at latitudes aligned parallel to the rotation axis as guided by the theory of magnetic flux rise in rapidly rotating stars \citep{Schu1996, Granzer2000, Isik2011, isik2018}. This trend is also observed in global convection simulations, where the Coriolis force deflects rising flux tubes toward high latitudes, leading to polar spot formation and altered magnetic field symmetry \citep{Brown08, viv18}.

\begin{figure}
    \centering
    \includegraphics[width=\columnwidth]{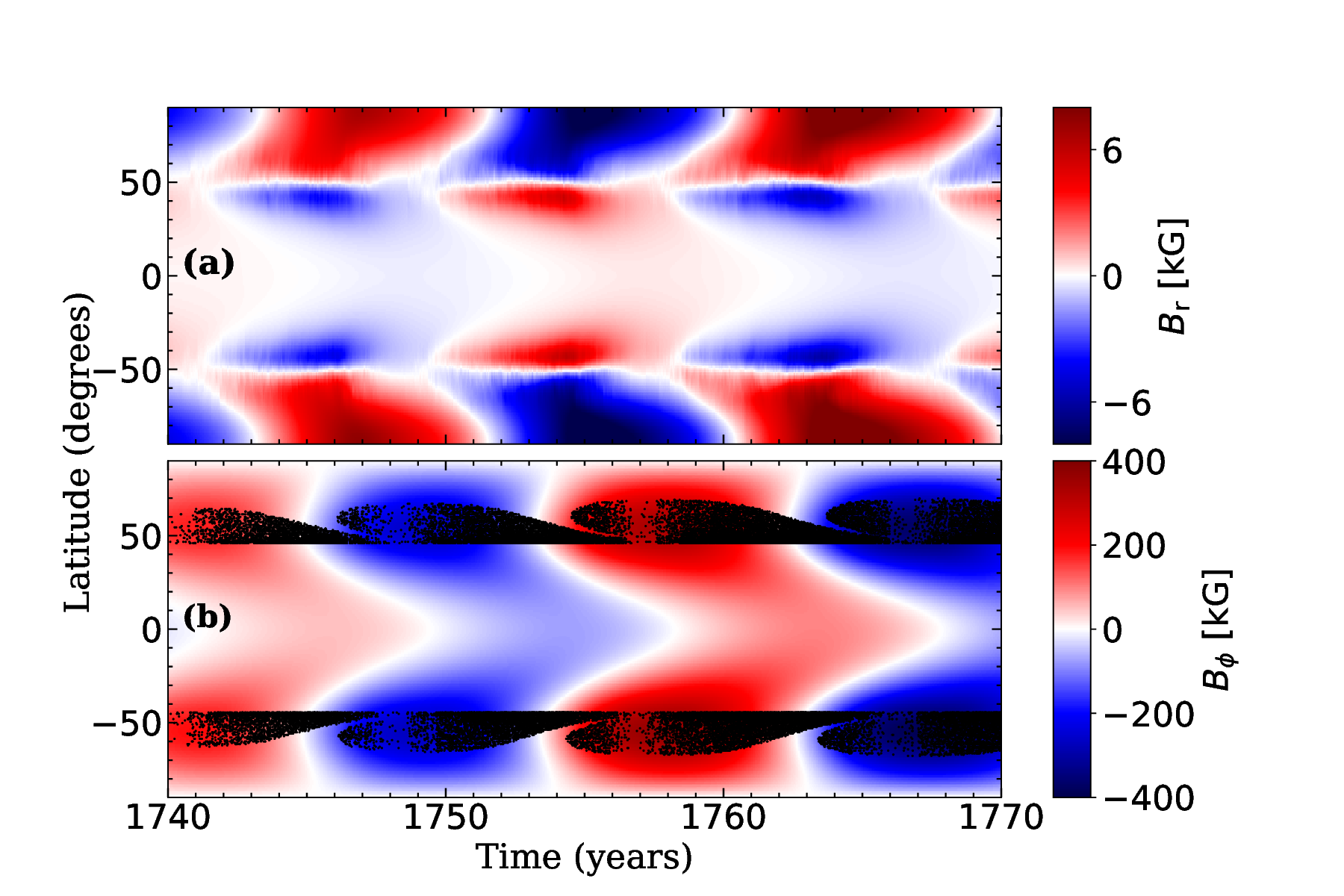}
    \caption {Time-latitude distribution of the (a) surface radial magnetic field $B_{\rm r}$ [in kG], and (b) toroidal field along with starspot distribution (black dots) for a star of 1 day rotation period for Case II.} 
    \label{fig:rot1p}
\end{figure}

As shown in \Fig{fig:rot1p}, all BMRs in this case appear above approximately $50^\circ$ latitude, and these high-latitude spots are inefficient for the cross-equatorial cancellation. The leading polarity flux cannot cancel the opposite polarity one from the other hemisphere, resulting in a predominantly quadrupolar magnetic field. 

\begin{figure}
    \centering
    \includegraphics[width=\columnwidth]{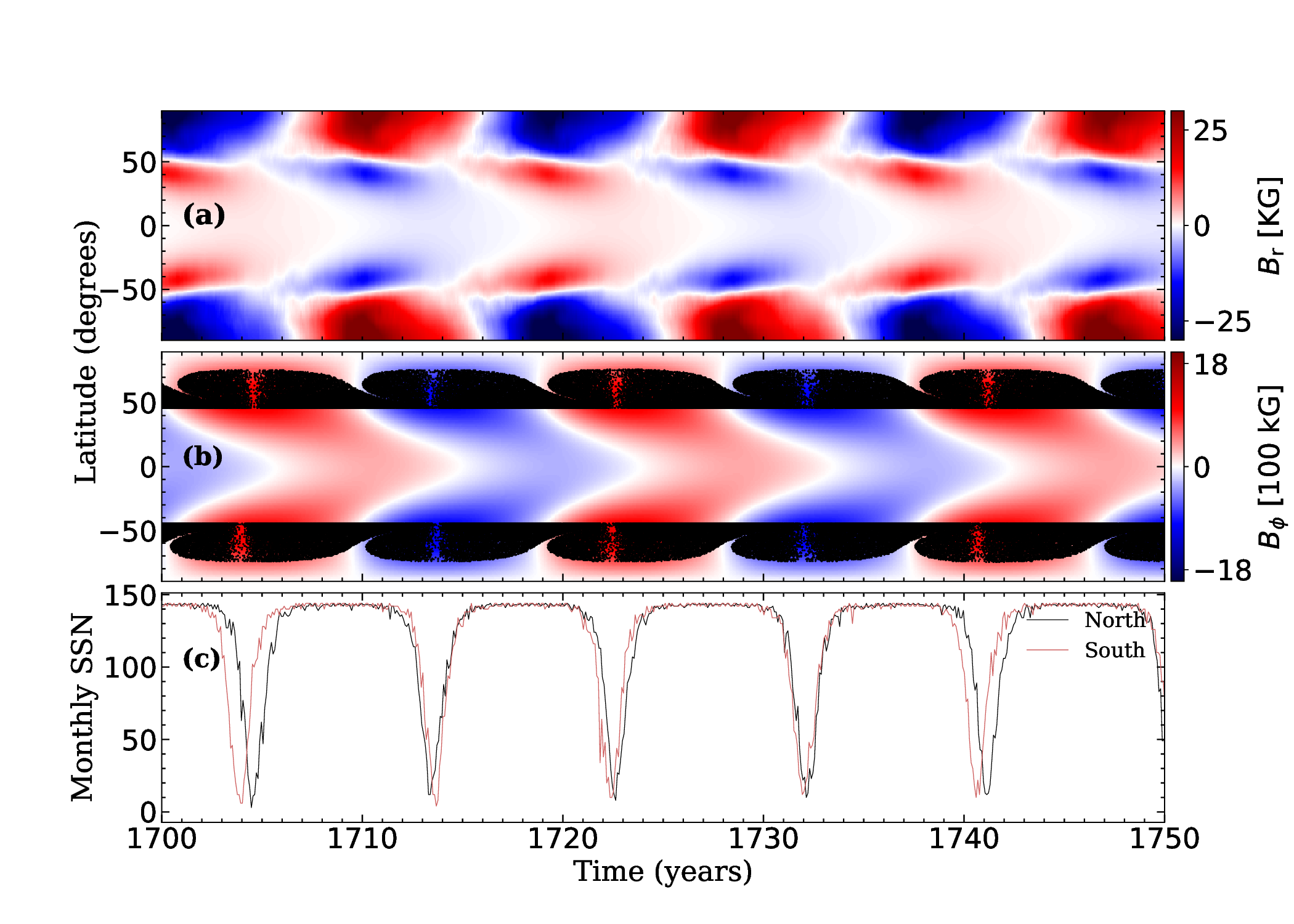}
    \caption{Same as \Fig{fig:rot1r}, but  for Case III.} 
    \label{fig:rot1pj}
\end{figure}

In Case III, where the BMR tilt is scaled by the rotation rate of the star 
(\Eq{eq:joyscalling}), we observe a strong increase in the magnetic field strength. The increase in field is due to the increase in the tilt angle of BMR. The magnetic field polarity is still remains quadrupolar (\Fig{fig:rot1pj}a) 
due to the high latitudes of spot emergence. Additionally, the spot eruption rate is quite high in this case. This is because the time delay (between two successive spots) is regulated by the magnetic field---with the increase of toroidal field, the delay distribution becomes narrow; see \Eq{eq:delay}. However, the delay cannot become smaller than the numerical time step, and thus, the monthly number of spots cannot exceed a certain value during the magnetic maximum; see \Fig{fig:rot1pj}(b).

\begin{figure}
     \includegraphics[width=1\columnwidth]{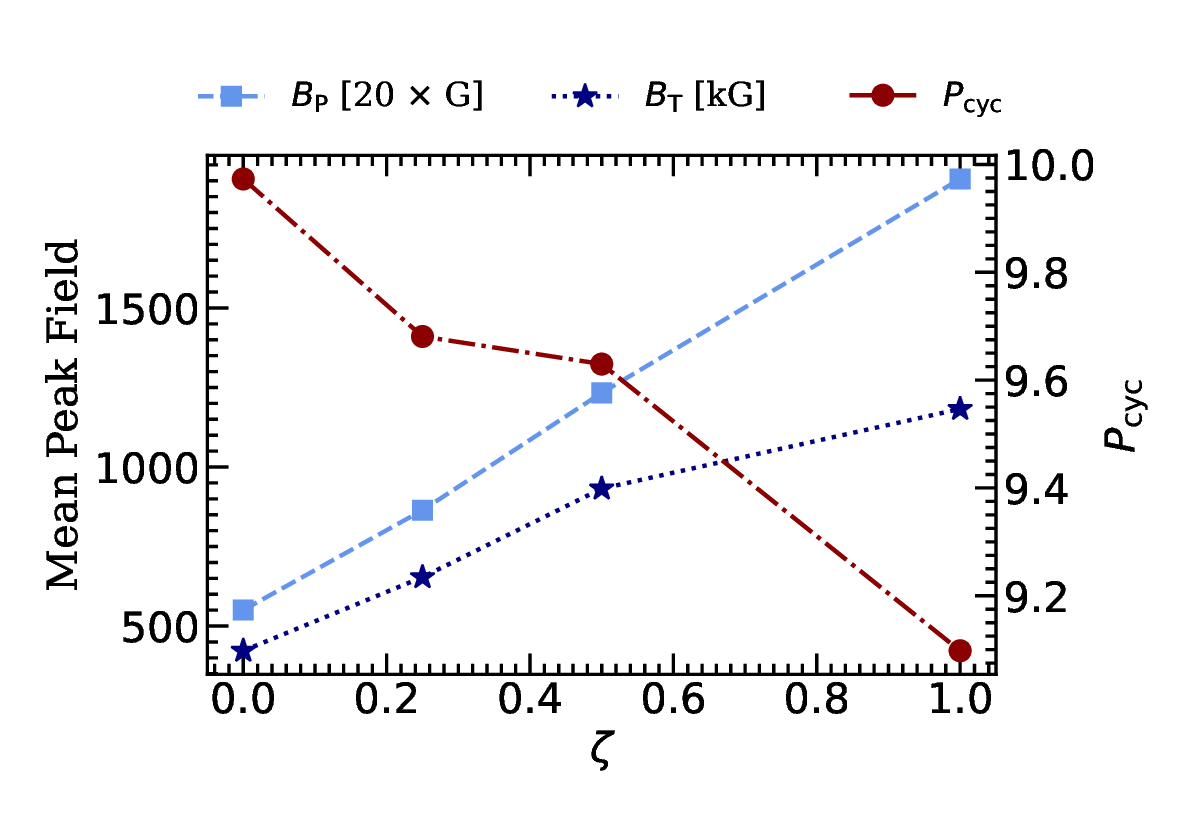}
        \caption{ The variation of the mean peak field for a rotation period (in years) of 1 day with the $\zeta$, the factor that shows the dependency of the rotation period on the tilt angle.  Here, $B_P$ and $B_T$ represent the poloidal and toroidal magnetic field components.}
    \label{fig:zetadep}
\end{figure}

Furthermore, to check the robustness of the dependence of the rotation rate on the amplitude of Joy's law, we used a sample rotation period of 1 day and varied the parameter $\zeta$ in the range of 0 to 1. As shown in \Fig{fig:zetadep}, we observe that the magnetic field strength increases with greater dependence of rotation period on Joy's law. Additionally, the cycle period becomes shorter with increasing dependency on rotation rate. This is because, with the increase of $\zeta$, the magnetic field gets stronger, and the tilt angle increases. Together, these make the dynamo more efficient at generating the poloidal field, which helps it flip the old magnetic polarity faster, leading to a shorter cycle period.

\begin{figure}
     \includegraphics[width=1.05\columnwidth]{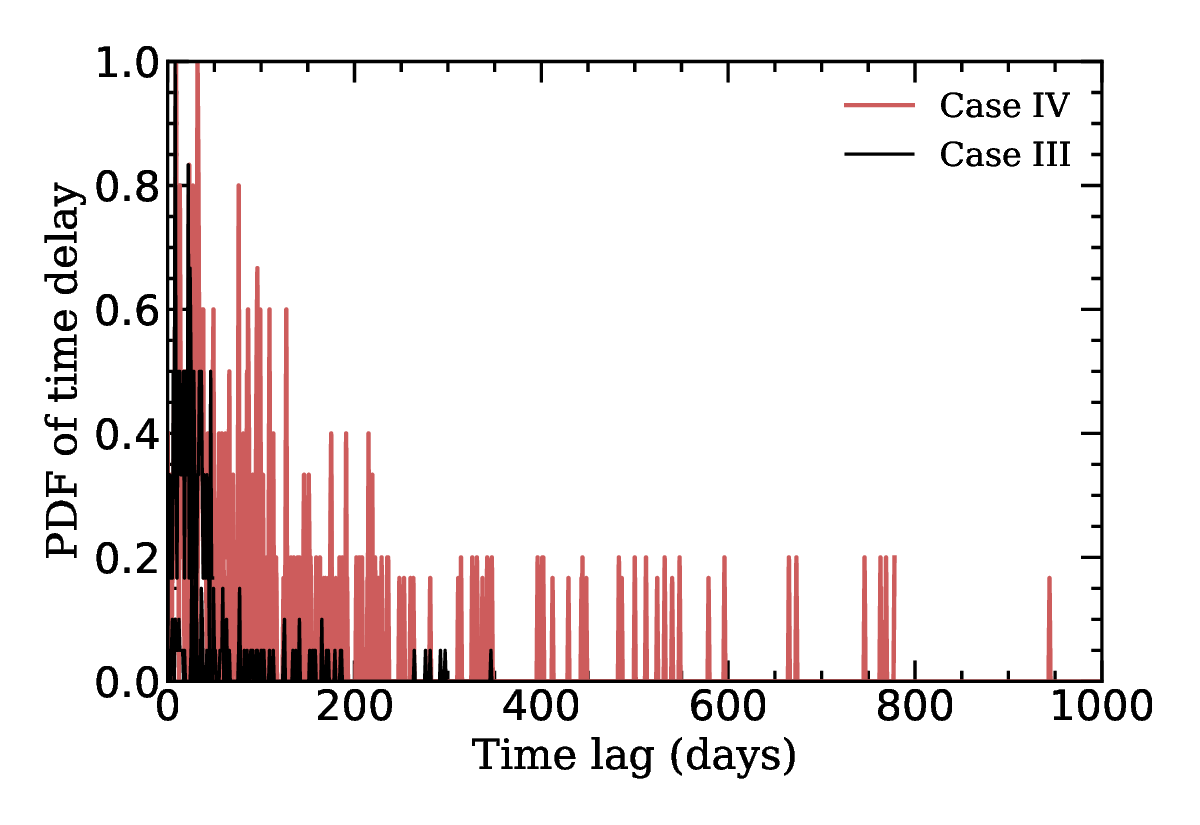}
     \caption{ Distribution of the time delays (lags between the successive BMR emergences) obtained from Cases III and IV for 1 day rotating star.}
    \label{fig:delayplot}
\end{figure}

\begin{figure}
    \centering
    \includegraphics[width=\columnwidth]{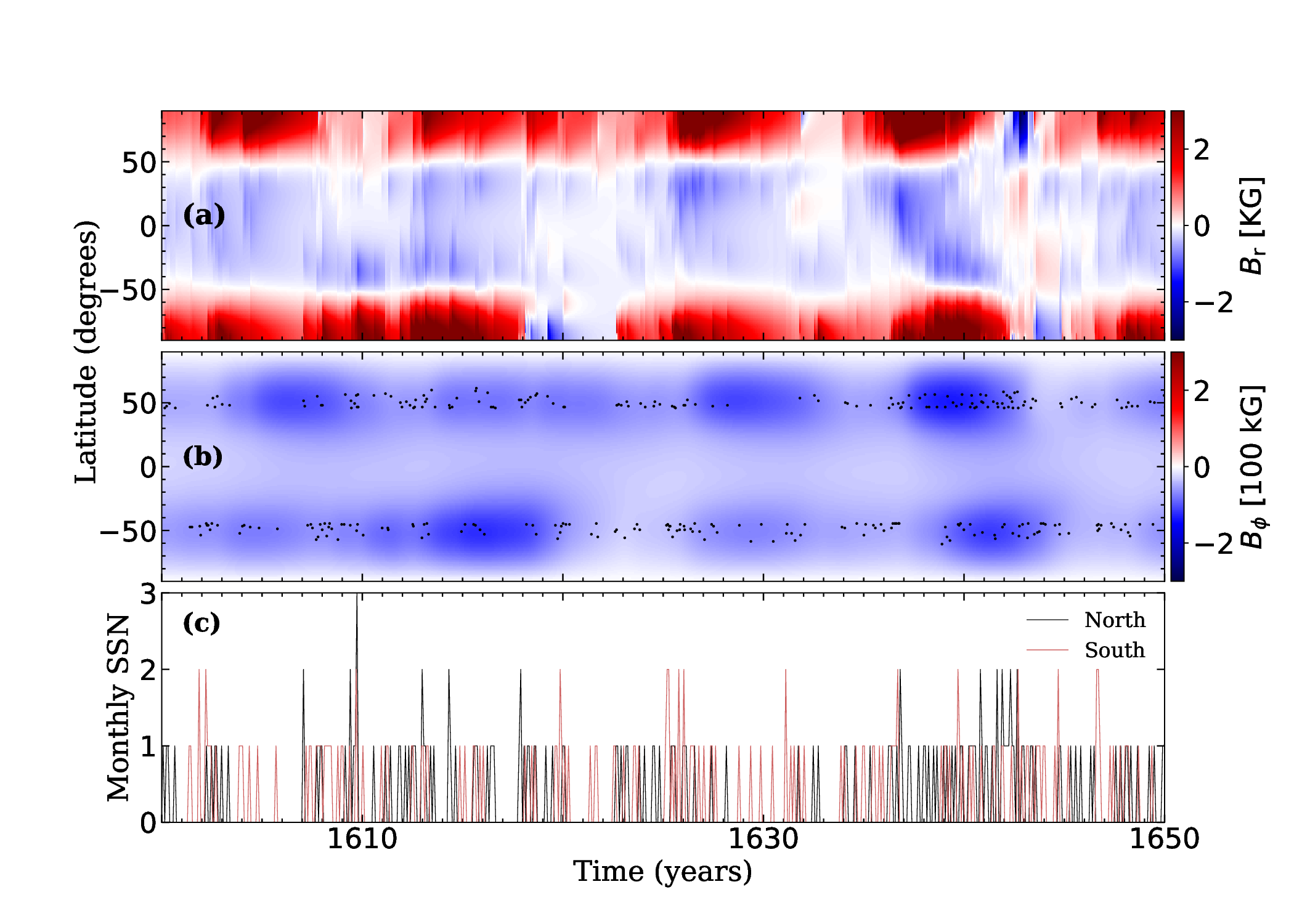}
    \caption{Same as \Fig{fig:rot1r}, but  for Case IV} 
    \label{fig:rot1delpar}
\end{figure}

Lastly, Case IV reveals an interesting trend wherein both the time delay between spot emergences and the size of the starspots are increased for rapidly rotating stars. The time delay distribution for this case, in comparison to Case III, is shown in \Fig{fig:delayplot}, clearly showing the increased in Case IV as compared to the other cases. Additionally, as indicated by \Eq{eq:flux}, the magnetic flux associated with spots also increases in this case. The flux distribution shifts toward higher values—approximately two orders of magnitude greater than in the previous cases, where the flux distribution was considered to be the same as that of the Sun.
As shown in \Fig{fig:rot1delpar}, we do not observe consistent polarity reversals or a well-defined magnetic cycle, in this case. However, the dynamo remains active, sustaining a non-zero magnetic field throughout the simulation. The absence of regular cycles is likely due to the reduced frequency of BMR emergence, which disrupts the balance required for systematic polarity reversals. Such non-cyclic but persistent dynamo states have also been seen in global dynamo simulations, particularly in the presence of strong rotational constraints and reduced convective mixing \citep{Nel13, viv19, brun22}.
Therefore, in Case IV, the BL dynamo operates only partially, maintaining non-zero fields without clear cycles.

\subsection{Stars of different rotation periods}

\begin{figure}
    \centering
    \includegraphics[width=1.05\columnwidth]{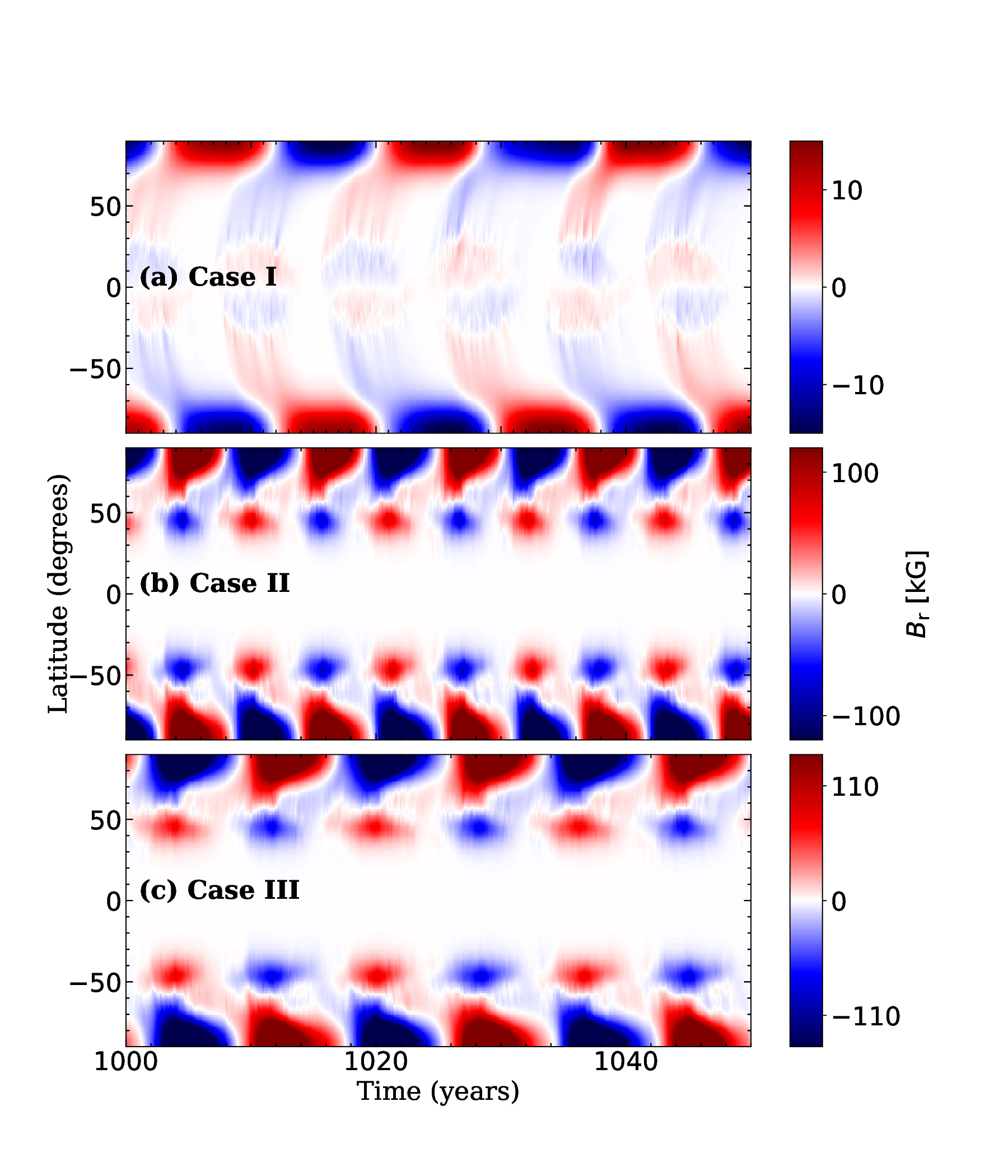}
    \caption{Time-latitude distribution of the surface radial magnetic field $B_{\rm r}$ [in kG] for a star of 10 days rotation period for Cases I--III.} 
    \label{fig:rot10d}
\end{figure}

We now consider the results of stars at different rotation rates. In Case I, for which spots are deposited in radial direction, i.e., low-latitude eruptions, we observe regular polarity reversals and smooth cycles with predominantly dipolar magnetic field for all the stars. As a representative example, the distribution of the surface radial field for the 10-day rotating star is shown in \Fig{fig:rot10d}(a). 

\begin{figure}
    \centering
    \includegraphics[width=1.1\columnwidth]{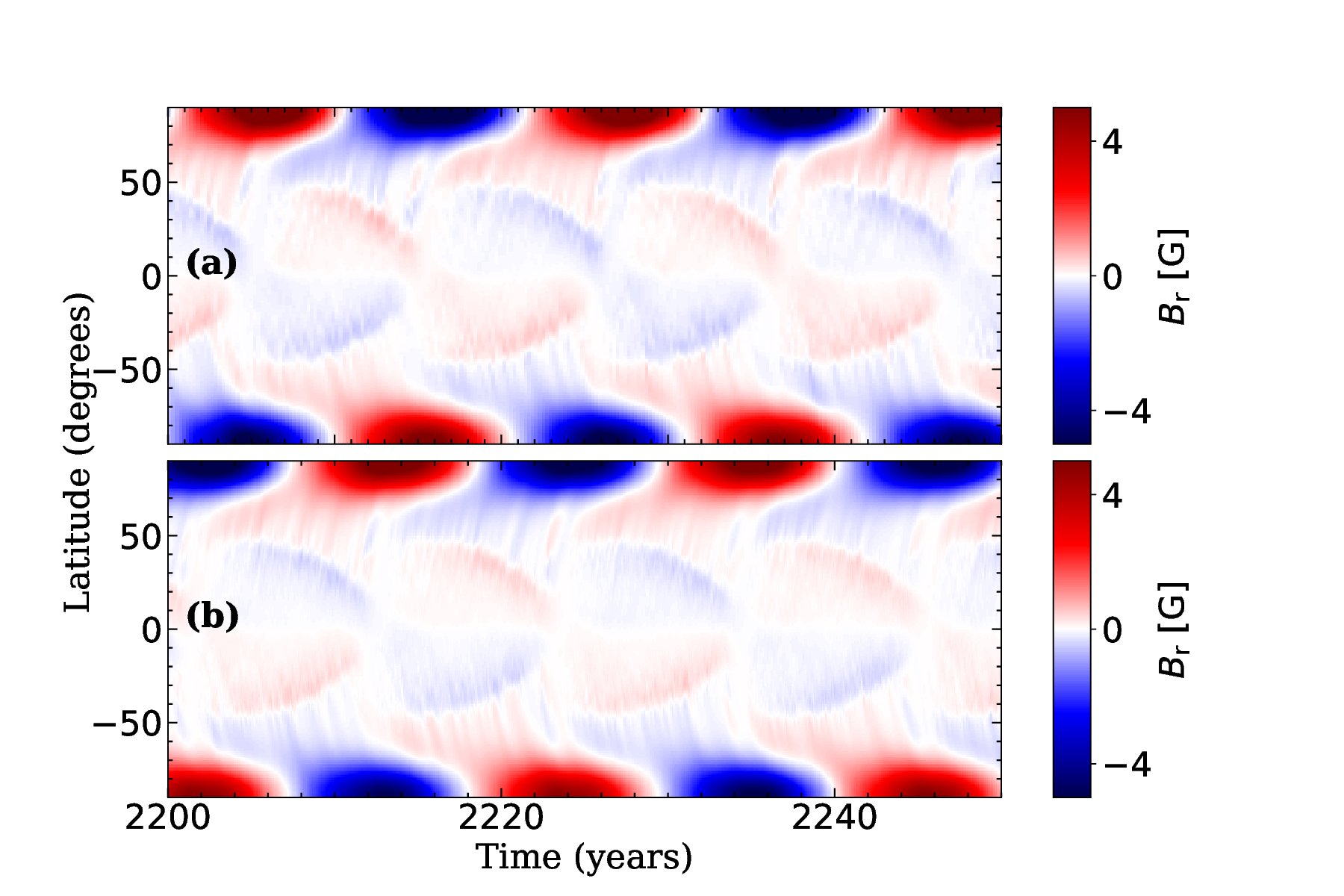}
    \caption{Time-latitude distribution of the surface radial magnetic field $B_{\rm r}$ [in kG] for a star of 30 days rotation period for Case I $\&$ III (Note: For stars with rotation periods $\geq 15$ days, Case III becomes identical to Case II).} 
    \label{fig:rot30d}
\end{figure}

In both Cases II and III, we also observe a magnetic cycle with regular polarity reversal. But again, because of the avoidance of low-latitude spots (parallel rise), we again observe a quadrupolar magnetic field for all stars with rotation period up to 10 days. A representative magnetic field distribution for the 10-day rotating star is shown in \Fig{fig:rot10d}(b) and (c), respectively, for Cases II and III. 
The major difference between these two cases is that the magnetic field in Case III is stronger as the tilt angle is scaled up by the rotation rate.  With the increase of rotation period from 10 days to 15 days, as the spot eruption zone moves to low latitude, which causes a decrease in the tilt, the magnetic field strength drops, and it continues to drop till the lowest considered case of 30-day rotation period. The field distributions for rotation period 30-day for Cases I and III are shown in \Fig{fig:rot30d}. Here we again recover regular solar-like oscillation with dipolar parity of magnetic field. Cases I and III display almost similar solutions because the only difference in Case III is that the amplitude of Joy's law is by a factor of 25.38/30. Further discussion on the dependence of magnetic field strength on rotation rate follows below.

We note that the polarity of the magnetic field is independent of the initial magnetic configuration.  In Case III (parallel rise) for fast rotators, we find that even if the simulations are initialised with dipolar parity, the magnetic field configuration flips to a quadrupolar one. Similarly, in Case I (radial rise/low latitude spots), even when we initialise the simulation with a quadrupolar field, it flips to dipolar.
Furthermore, we do not observe any change in parity during the course of the simulation in any star at the steady state, whereas in the solar dynamo model, this has been observed due to fluctuations in the dynamo parameter \citep{HN19}, which we kept fixed in each run.

Interestingly, in previous axisymmetric dynamo simulations \citep{Hazra19a, Vindya23} with $\alpha$ parametrization for the BL process for stars with rotation period less than 10 days also produced quadrupolar magnetic field. In our 3D dynamo model, in Case I, we always get dipolar parity because of the low-latitude eruption.

\begin{figure}
    \includegraphics[width=\columnwidth]{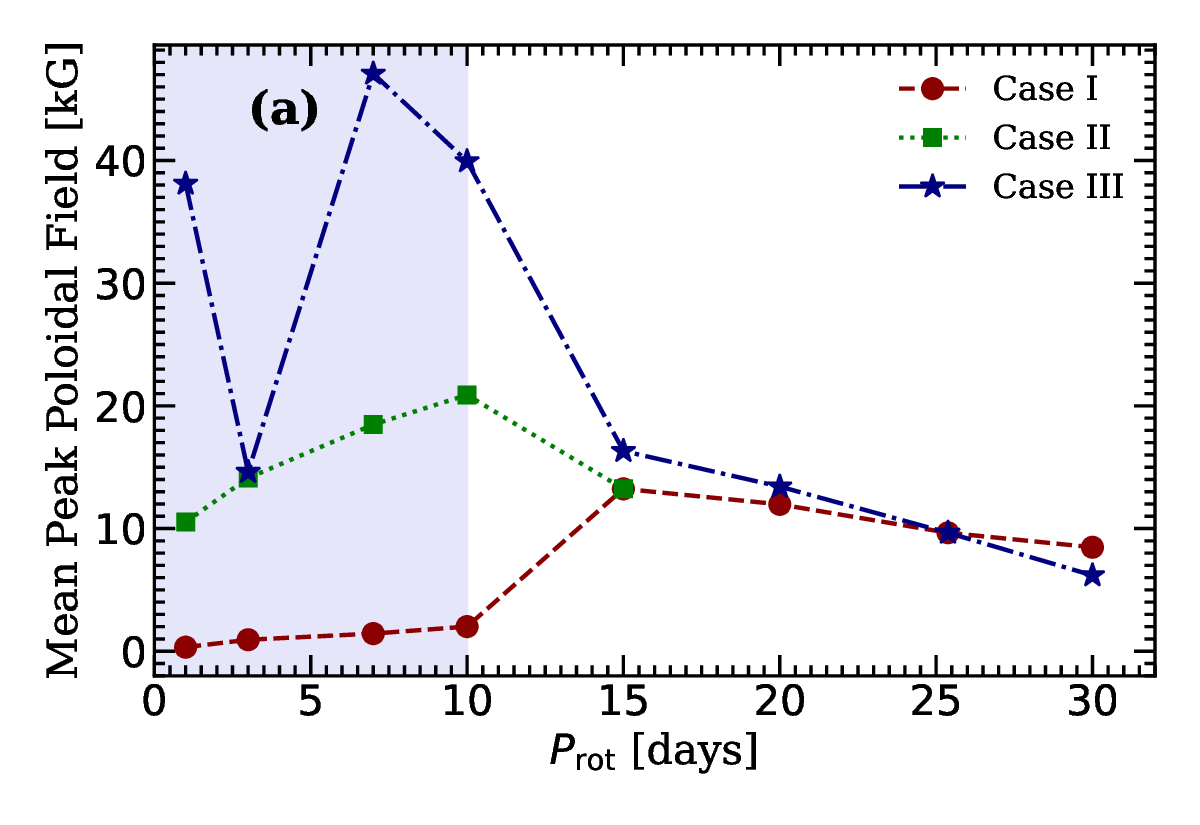}
    \includegraphics[width=\columnwidth]{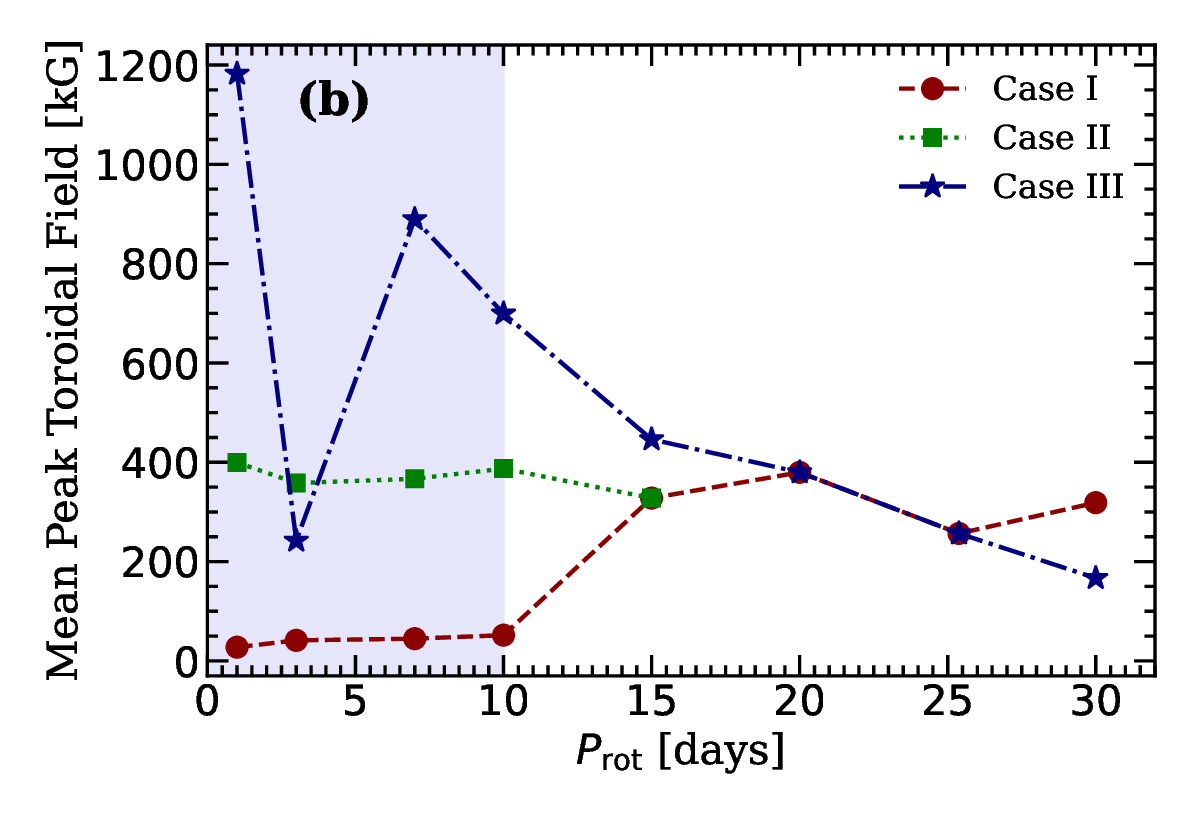}
    \caption{Variation of the mean peak (a) toroidal and (b) poloidal magnetic fields with the rotation period of the star for different cases of simulations. The shaded area represents the range of rotation periods for which the spots are deposited in parallel to the rotation axis, i.e., void of low-latitude eruptions.  
     }
    \label{fig:pf_tf}
\end{figure}

The variation of the toroidal and poloidal magnetic field strengths as function of the rotation period is shown in \Fig{fig:pf_tf}. For Case I (red points), both fields slowly increase with the increase of rotation period up to about 10 days and decrease beyond that point. 
The trend is due to the fact that as a star rotates faster, the shear decreases, and the omega effect decreases as well \citep{KR99, KO12b} (we confirmed this decreasing trend after computing $\Delta \Omega$ for each star). At this point, the meridional flow becomes a significant factor in determining the trend of the poloidal and toroidal fields with the rotation period, as recently demonstrated in \citet{Vindya24}.
For fast rotators, starting with a rotation period of 1 day, the surface meridional flow is very strong, while the flow within the bulk of the CZ is negligible \citep[see Figure 7 of][]{Vindya24}, both contributing to an extremely weak polar field \citep[also see Figure 8a of][]{Vindya24}. 
As the rotation period increases, the meridional flow within the bulk strengthens, leading to an increase in the polar field.
Conversely, at a rotation period of 30 days, the bulk flow is strongest, but the surface flow is weakest, resulting in a weak polar field again. When the rotation period decreases from this point, the surface flow intensifies, contributing to a stronger polar field.
Thus, the polar field increases when transitioning from low to high meridional flow in the bulk or from low to high surface flow. We now turn to the variation of magnetic field strengths in the other models.

The magnetic field strengths for Cases II and III, as functions of rotation period are shown by green and blue points in \Fig{fig:pf_tf}.  
For Case III, in all the stars, $\zeta$ in \Eq{eq:joyscalling} is taken as 1. 
For Case II, where the BMRs rise parallel to the rotation axis in rapidly rotating stars, we observe a similar increasing trend in the poloidal field as in Case I, whereas the toroidal field decreases a bit at rotation period 3 days and increases thereafter.
This decrease in the magnetic field with the rotation rate is again due to the reduction in the variation of angular velocity and the decrease in the meridional flow within the bulk of CZ with the increase in the rotation rate of the star.

We note that Case I and Case II become identical for stars of rotation period $\geq 15$ days.
Therefore, the behavior for Case II is interesting because the activity initially increases with the increase of rotation rate (from right to left in \Fig{fig:pf_tf}) and then it decreases.  This behavior is somewhat consistent with the stellar observation of coronal and chromospheric emission vs rotation rate \citep{Noyes84a, wright16} and previous results from axisymmetric dynamos \citep{KKC14, Hazra19a, Vindya23}. 
Case III shows somewhat similar behavior except that the activity level increases rapidly with the decrease of rotation period. This is because in this case, the tilt increases with the decrease of rotation period.
In the rapid rotator regime, the fields show a somewhat complex trend: they decrease after a rotation period of 1 day, then increase suddenly after a rotation period of 7 days, and decrease again. This decreasing trend remains consistent in the slow rotator regime.

\subsection{Cycle period vs rotation period}
\begin{figure}
    \centering
    \includegraphics[width=\columnwidth]{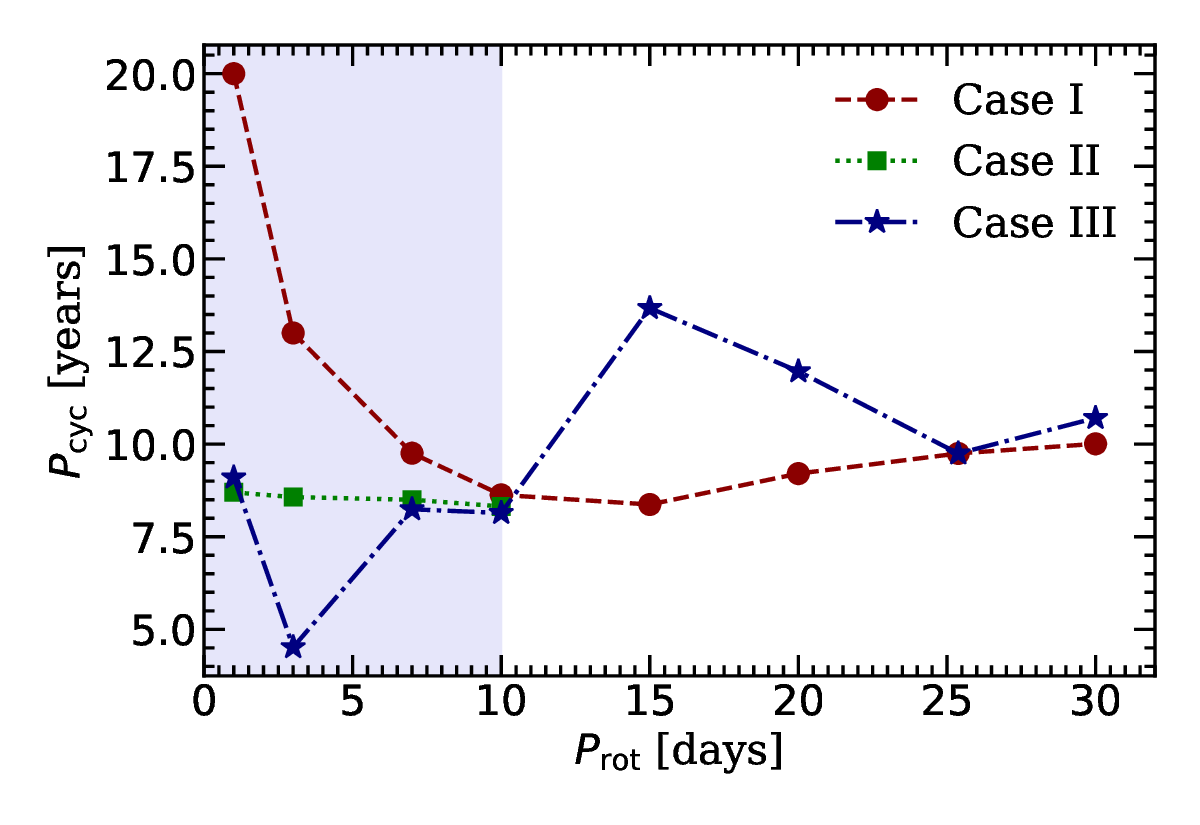}
    \caption{Variation of the cycle period with the rotation period of solar-type stars having rotation periods varying from 1--30 days.} 
    \label{fig:pcy}
\end{figure}

We now analyze the cycle duration of stars for all cases, and the trends are shown in \Fig{fig:pcy}.
In Case I, we observe a monotonic decrease in the cycle period with the rotation period for fast rotators. This happens because, as the rotation period decreases (or rotation rate increases), the meridional flow becomes weaker (although the flow speed increases in the thin layers near the top and bottom boundaries). 
However, after a rotation period of 15 days, there is a slight increase in the cycle period at higher rotation periods. The slight increase in the cycle period (after a rotation period of 15 days) is because the generation of the poloidal field weakens as the star spins down, and the poloidal field needs more time to reverse the old field.
This monotonic decrease in cycle duration for rapidly rotating stars is also observed in Case II. 
The result is compatible with global simulations of \citet{Gue19}, kinematic and no-kinematic simulations of \citep{pipin21}, and most importantly with the observational results of chromospheric and photometric studies of the solar-type stars \citep{BoroSaikia18}. These limited observations seem to show a rapid increase in the cycle period with the increase of the rotation rate for fast-rotating stars, which is consistent with the trend found in our Cases I--II. 
Meanwhile, the observed data for slow rotators show an increasing trend of activity cycle period with an increase in rotation period, which is somewhat in agreement with our Case I. However, for Case III, the trend is quite complicated; increasing first till the rotation period of 15 days and declines afterward.

\section{Conclusions and Discussions} \label{sec:conclusion}

In this study, we use the 3D STABLE (Surface flux Transport And Babcock–Leighton) dynamo model, which realistically captures the generation of poloidal field---the decay and dispersal of star spots to explore the operation of BL dynamo in solar-type stars. In particular, we address the possibility of BL dynamo in rapidly rotating stars of rotation period 1 day and more, for which star spots are predominantly formed at high latitudes. 
In our study, we consider $1 M_\odot$ mass stars of different rotation periods, starting from 1 day to 30 days.
Meridional flow and differential rotation profiles are derived from a mean-field hydrodynamics model tailored for stars with different rotation periods. These profiles are incorporated into the dynamo model to investigate the operation of the Babcock–Leighton dynamo under various scenarios. 

We find consistent results of increasing magnetic field strength with the decrease of rotation period and a tendency of saturation (or even decrease) of magnetic field for stars with rotation period less than 10 days, in qualitative agreement with observations \citep{Noyes84a, wright11}.

Additionally, we find cyclic magnetic fields in all cases except case IV of the 1-day rotating star, for which the magnetic field is irregular.
Overall, our findings show that the BL dynamo efficiently operates with regular polarity reversals and predominantly dipolar parity across all stars considered, except for rapidly rotating stars with rotation periods $\leq 10$ days. For stars with rotation periods $\leq 10$ days, when the toroidal flux rises parallel to the rotation axis, spots appear at high latitudes, leading to a low-latitude zone of avoidance. These high-latitude spots lead to inefficient cross-equatorial cancellation of the leading polarity flux across the equator, thereby maintaining a quadrupolar magnetic field, in contrast to the dipolar field as obtained in the radial rise scenario with low-latitude spots. Under these conditions, the mechanism becomes geometry-modified, resulting in quadrupolar parity or irregular cycles. Future work incorporating nonlinear effects and additional dynamo ingredients could further clarify this behavior.

While several observations and theoretical models support the predominance of high-latitude spots in rapidly rotating solar-type stars (driven by strong Coriolis forces and poleward-propagating dynamo waves) \citep{Schu1992, KO15, Kap17}, some stars can show low-latitude spot emergence also, indicating complex dynamo behavior. For instance, Doppler imaging of AB Doradus ($P_{\mathrm{rot}} = 0.51$ days) reveals occasional low-latitude spots (~$20^\circ - 30^\circ$) alongside dominant polar spots, likely due to transient flux emergence in a distributed dynamo \citep{donati2003}. Similarly, LO Pegasi ($P_{\mathrm{rot}} = 0.42$ days) and HD 171488 ($ P_{\mathrm{rot}} = 1.34$  days) show low-latitude spots (~$10^\circ - 40^\circ$) in some epochs, attributed to mixed dynamo modes combining solar-like and rapid-rotator characteristics \citep{barnes05}. EK Draconis ($P_{\mathrm{rot}} = 2.5$ days) also displays low-latitude spots (~$15^\circ - 35^\circ$), reflecting a dynamo retaining solar-like features \citep{Stras98}. These examples highlight that, while high-latitude spots dominate, low-latitude spots may occur in rapidly rotating stars. The reason could be that even though rapid stellar rotation enhances the Coriolis force, which tends to deflect rising magnetic flux tubes toward the poles, sufficiently strong toroidal magnetic fields can overcome this effect. If the flux tubes rise rapidly—due to enhanced buoyancy from stronger fields or a more superadiabatic stratification in the upper convection zone—they experience less poleward deflection and can still emerge near the equator \citep{isik24}. 

In this study, we chose to focus exclusively on high-latitude starspots, as they are the dominant feature in rapidly rotating stars—mainly due to stronger Coriolis forces and poleward transport of magnetic flux \citep{KO15}. By narrowing our scope to this regime, we were able to capture the key magnetic characteristics linked to rapid rotation and assess how well the Babcock–Leighton process functions under these conditions, which was the main goal of our work. In the future, it would be interesting to explore low-latitude spot formation as well. However, including those features would require assumptions about low-latitude spot distributions that are not yet well constrained by observations and would make the model significantly more complex. Even so, this remains a promising direction for future work.

\section*{Acknowledgement}
The authors thank the anonymous referee for providing useful comments to improve the presentation of the manuscript. 
The authors also acknowledge the computational support and the resources provided by the PARAM SHIVAY Facility under the National Supercomputing Mission, the Government of India, at IIT (BHU) Varanasi. B.B.K. acknowledges the  Anusandhan National Research Foundation (ANRF) for providing financial support through the MATRIC program (file no. MTR/2023/000670).

\section*{Data Availability}
The dynamo simulations in this study are carried out using the code STABLE \citep{MD14, MT16} developed at the National Center for Atmospheric Research. The simulation data and analysis scripts used in this article can be made available upon reasonable request.

\bibliographystyle{mnras}
\bibliography{paper}{}

\bsp	
\label{lastpage}
\end{document}